\documentclass[reprint, amsmath, amssymb, aps]{revtex4-1}
\usepackage{xcolor}
\usepackage{comment}
\usepackage{tikz, adjustbox}
\usepackage{textcomp}
\usetikzlibrary{arrows.meta, bending}
\newcommand{\nicesection}[1]{
\vspace{0em}
\begin{center}
\noindent\textbf{#1}
\end{center}
\vspace{0em}
}
\usepackage{microtype}

\usepackage{soul}

\usepackage[colorlinks=true, linkcolor=blue, citecolor=blue, urlcolor=blue]{hyperref}

\DeclareMathOperator{\Tr}{Tr}
\DeclareMathOperator{\Pu}{P}

\def\dd{\text{d}}
\newcommand{\ba}{\bar{A}}
\newcommand{\bp}{\bar{\partial}}

\allowdisplaybreaks

\begin{document}

\title{A Carroll Limit of AdS/CFT: A Triality with Flat Space Holography?}

\author{\bf{Andrea Fontanella}}
\email{andrea.fontanella[at]tcd.ie} 
\affiliation{\vspace{2mm} School of Mathematics $\&$ Hamilton Mathematics Institute, 
Trinity College Dublin, Ireland}

\author{\bf{Oliver Payne}}
\email{payneo[at]tcd.ie} \email{oliverbgpayne[at]gmail.com}
\affiliation{\vspace{2mm} School of Mathematics $\&$ Hamilton Mathematics Institute, 
Trinity College Dublin, Ireland}

\begin{abstract}

We construct a novel holographic duality by taking a Carrollian limit of the AdS/CFT correspondence, relating string theory in a Carroll bulk geometry to a Carroll $\mathcal{N}=4$ Super Yang-Mills theory. We further propose the existence of an underlying triality connecting Carroll string theory, relativistic string theory in flat spacetime, and Carroll gauge theory. Finally, we analyse the symmetries of the Carroll gauge theory, showing that they are non-linearly realised and do not close under ordinary Lie brackets.

\end{abstract}

\maketitle

\begin{figure*}
\centering
\begin{adjustbox}{width=.7\textwidth}
\hspace{-8mm}
\begin{tikzpicture}[
    every node/.style={
        draw, 
        minimum width=1.2cm, 
        minimum height=0.8cm, 
        font=\scriptsize\rmfamily,
        rounded corners=2pt, 
        align=center
    },
    >=Stealth
]

    \node (A) at (2,4) {ST in AdS$_5 \times$ S$^5$};
    \node (B) at (5.2,4) {$\mathcal{N}=4$ SYM};
    \node (C) at (2,2) {ST in Carroll \\AdS$_5 \times$ S$^5$};
    \node (D) at (5.2,2) {Carroll\\ $\mathcal{N}=4$ SYM};
    \node (E) at (-1,1.6) {ST in Mink$_{10}$};
    \node (F) at (5.2,0) {ST in Mink$_5\times X^5$};

    \draw[<->, double, double distance=1pt, >=Stealth, red] (A) -- (B);
    \draw[<->, double, double distance=1pt, >=Stealth, blue] (C) -- (D);

    \draw[-{>[length=1.3mm]}] (A) -- node[pos=0.5, font=\tiny, draw=none, inner sep=0pt, xshift=11pt] {$c \to 0$} (C);
    \draw[-{>[length=1.3mm]}] (B) -- node[pos=0.5, font=\tiny, draw=none, inner sep=0pt, xshift=11pt] {$c \to 0$} (D);

    \draw[-{>[length=1.3mm]}] 
        (A) -- node[pos=0.65, above, yshift=2pt, font=\tiny, draw=none, inner sep=0pt] {$R \to \infty$} (E);

    \draw[-{>[length=1.3mm]}, bend right=35] 
        (E) to node[pos=0.5, above, yshift=-5pt, xshift=25pt, font=\tiny, draw=none, inner sep=0pt] {compactification} (F.west);

    \draw[<->, double, double distance=1pt, >=Stealth, blue] (D) -- (F);
    \draw[<->, double, double distance=1pt, >=Stealth, blue] (C) -- (F);

\end{tikzpicture}

\end{adjustbox}
\caption{The red double arrow indicates Maldacena's holography. The blue double arrows represent the proposed triality. The acronyms ST and SYM stand, respectively, for String Theory and Super Yang-Mills.}
\label{fig:triality}
\end{figure*}
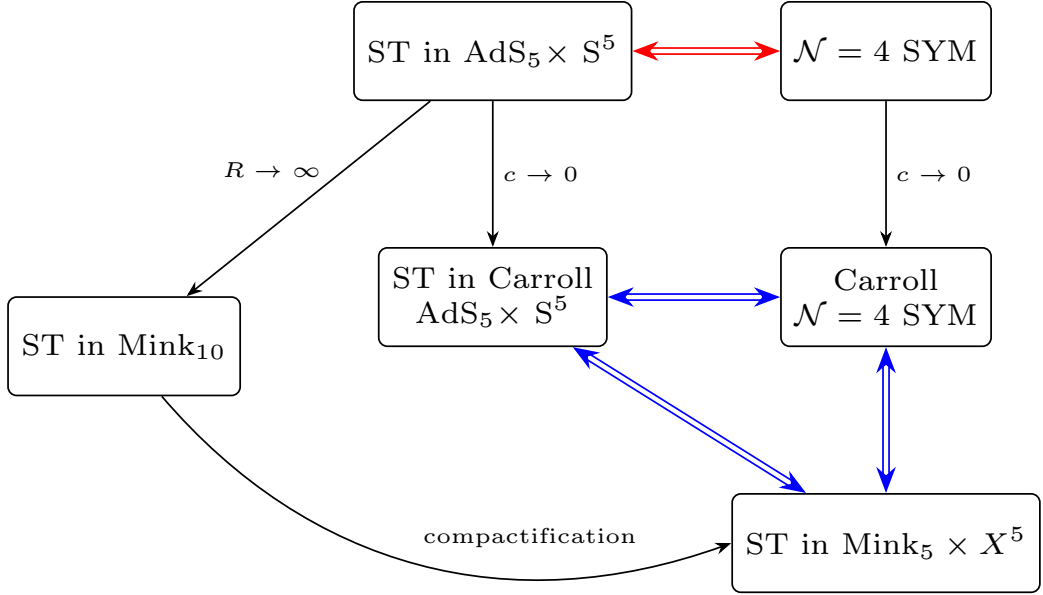


\nicesection{Introduction}

The study of holography in high-energy physics, namely a duality between a theory of gravity and a theory without gravity defined on the boundary of spacetime, is becoming increasingly popular due to its first-principle appeal, but also for its potential as a tool in defining quantum gravity \cite{Takayanagi:2025ula}. 

The first concrete example of holography was given by Maldacena \cite{Maldacena:1997re} as an equivalence between string theory in AdS$_5\times$S$^5$ and $\mathcal{N} = 4$ Super Yang-Mills (SYM) in 4d flat spacetime. The setting considered by Maldacena is type IIB string theory in flat spacetime with a stack of D3-branes. At strong string coupling, the system is described by closed strings probing a black D3-brane geometry. At weak string coupling, it is instead described by interacting closed and open strings. Maldacena's idea is that these two pictures are equivalent at low energy, giving birth to the celebrated duality. 

In recent years, there has been a lot of interest in exploring whether holography still applies to spacetimes that are not Anti-de Sitter. For example, formulating holography in flat spacetime is currently of high interest. At present, flat space holography admits two (equivalent) formulations: the dual field theory is either defined on the celestial sphere of co-dimension 2 \cite{Pasterski:2016qvg, Pasterski:2021raf, Donnay:2023mrd, Raclariu:2021zjz, McLoughlin:2022ljp}, or on a Carroll (i.e. non-Lorentzian) manifold of co-dimension 1 \cite{Donnay:2022aba, Donnay:2022wvx, Bagchi:2016bcd, Bagchi:2023cen, Bagchi:2025vri, Nguyen:2023vfz, Nguyen:2023miw}. 
The ``Carrollian'' approach is based on the observation that the asymptotic symmetries of the flat spacetime are given by the Bondi, van der Burg, Metzner and Sachs (BMS) group, isomorphic to a conformal extension of the Carroll group \cite{Duval:2014uva}.


\nicesection{A triality with flat space holography?}

\paragraph*{Statement of a problem.}
A very legitimate question is whether one can derive flat space holography via a limit of the AdS$_5$/CFT$_4$ correspondence, see \cite{Hijano:2019qmi, Ciambelli:2018wre, Campoleoni:2023fug, Li:2021snj, Hijano:2020szl, Alday:2024yyj, Lipstein:2025jfj, Kulkarni:2025qcx, Kraus:2024gso, Duary:2022pyv, Fareghbal:2024lqa, Duary:2024kxl, Marotta:2024sce, Surubaru:2025fmg, Hao:2025btl, Arenas-Henriquez:2025rpt, Poulias:2025eck, Neuenfeld:2025wnl, Grumiller:2023rzn} for recent developments. In the bulk theory, this is achieved by a large radius limit. Since the AdS$_5$ and S$^5$ spaces must have equal radius, the large radius limit will completely flatten the spacetime, leading to Mink$_{10}$. 
On the other hand, in the gauge theory side the large radius limit effectively acts as a Carroll limit, as elucidated in \cite{Alday:2024yyj, Lipstein:2025jfj}, leading to Carroll $\mathcal{N} = 4$ SYM. 

At this point, it is clear that type IIB string theory in Mink$_{10}$ \emph{cannot} be holographically dual to the Carroll limit of $\mathcal{N} = 4$ Super Yang-Mills, for the simple fact that:
\begin{eqnarray}
    \partial \text{Mink}_{10} \neq \text{Carroll}_4
\end{eqnarray}
i.e. the conformal boundary of 10d flat spacetime is \emph{not} the Carroll Newton-Cartan version of 4d flat spacetime.

\paragraph*{Possible resolutions.} We propose two possible solutions: (i) the duality holds only after a compactification of Mink$_{10}$ to Mink$_5 \times X^5$, with $X^5$ a locally flat 5d compact Euclidean space \footnote{This aligns with \cite{Lipstein:2025jfj}, where the correlation functions of the gauge theory in the Carroll limit do not match directly the amplitudes of the bulk theory in the large radius limit. To match, the bulk theory needs to be truncated to an effective lower dimensional theory.
For a somehow related recent discussion, see also \cite{Sen:2025oeq}}; or (ii) the bulk theory is also Carrollian.

In this letter, we develop the second approach. The idea is to incorporate the Carroll limit into Maldacena's AdS/CFT construction. 
For the case of the non-relativistic limit, there is evidence that holography survives after taking the limit on both sides of the AdS/CFT \cite{Fontanella:2024kyl, Fontanella:2024rvn, Fontanella:2024hgv, Lambert:2024uue, Lambert:2024yjk, Lambert:2024ncn} \footnote{The topic of non-relativistic holography is quite old, and it started with Lifshitz and Schr\text{\"o}dinger holography {\cite{Maldacena:2008wh, Guica:2010sw, Christensen:2013lma, Dobrev:2013kha, Taylor:2008tg, Taylor:2015glc}}, where the boundary theory is non-relativistic, but the bulk is still Lorentzian. An example of holography where the bulk is also non-relativistic was formulated for string theory with non-relativistic world-sheet {\cite{Harmark:2006ta, Harmark:2016cjq, Harmark:2017rpg}}. Only recently {\cite{Fontanella:2024kyl, Fontanella:2024rvn, Fontanella:2024hgv, Lambert:2024uue, Lambert:2024yjk, Lambert:2024ncn}}, holography was suggested for the ``Gomis-Ooguri'' non-relativistic string with relativistic world-sheet, via taking limits of the AdS/CFT. After that, there has been a host of interesting work revolving around the topic of non-relativistic holography in the modern approach, see {\cite{Blair:2024aqz, Harmark:2025ikv, Blair:2025prd, Guijosa:2025mwh}}.}. It is natural to suggest that the same logic can be applied to the Carroll limit. 

The logic behind this approach is that one rescales fields on both string and gauge theories with a parameter $c$, and then take a limit. However, the main point is that the parameter $c$ is not necessarily the same parameter on both theories. The criteria are: (i) the gauge theory is defined on the Penrose conformal boundary of the string theory; (ii) the symmetries of the two theories match; and (iii) the non-Lorentzian limit commutes with the Maldacena's decoupling limit \footnote{These three conditions are part of a ``zeroth order test''. To confirm the duality, one should prove that observables from the two theories have the same expectation values.}. Broadly speaking, this method consists in looking for limits on both sides of the AdS/CFT that are sensibly related. 

\paragraph*{A triality?} In this letter, we incorporate the Carroll limit inside the Maldacena's construction of the AdS/CFT correspondence, leading us to propose a duality between string theory in the Carroll limit of the AdS$_5\times$S$^5$ geometry and the magnetic Carroll limit of $\mathcal{N} = 4$ SYM that appears from the DBI \footnote{Another instance where a magnetic Carroll gauge theory arises to be dual to bulk flat space in 2+1d was suggested in \cite{Cotler:2024cia}.}. Then, based on the above discussion regarding flat space holography, we expect that $\mathcal{N} = 4$ SYM in the Carroll limit is also dual to type IIB string theory on Mink$_5 \times X^5$.  It is tantalising to propose an underlying triality between Carroll string theory, string theory in flat spacetime and Carroll $\mathcal{N} = 4$ Super Yang-Mills. This idea unifies the web of holographic dualities, as summarised in fig. \ref{fig:triality}.


\newpage
\nicesection{Carroll limit in the gravity side}

We start by incorporating the Carroll limit into the ``gravity side'' of the Maldacena's holographic construction, requiring as a consistency condition that the near-horizon limit commutes with the Carroll limit. 
The starting point is the metric for a stack of $N$ D3-branes, given by 
\begin{eqnarray}
\label{D3_metric_z}
    \dd s^2_{\text{D3-brane}} &=& \frac{4 \pi  g_s N}{\sqrt{f(z)}} \left( -\dd t^2 + \dd x^i \dd x_i   \right) \\
    \notag
    &+& \alpha'^2 \sqrt{f(z)} \left(\frac{\dd z^2}{z^4} + \frac{1}{z^2} \dd \Omega^2_5 \right) \, , \\
    \notag
    \dd \Omega_5^2 &=& \left(\frac{4-y^2}{4+y^2}\right) \dd \phi^2 + \frac{16\ \dd y^2 }{\left( 4+y^2\right)^2} \, , \\
    \notag
f (z) &=& 1 + \frac{4 \pi  g_s N}{\alpha'^2} z^4 \, ,
\end{eqnarray}
where $g_s$ is the string coupling, $(t, x^i)$, with $i=1,2,3$, are coordinates along the world-volume of the D3-brane, and $\dd \Omega_5^2$ is the metric of the unit 5-sphere. We describe the 5-sphere metric $\dd \Omega_5^2$ in terms of Cartesian coordinates $(\phi, y^m)$, with $m=1,..., 4$ and $y^2 \equiv y^m y^n \delta_{mn}$.
The first step is to take the Maldacena's near-horizon limit $\alpha' \to 0$ in \eqref{D3_metric_z}, which gives us the famous AdS$_5\times$S$^5$ metric,
\begin{eqnarray}
\label{AdS5xS5_metric}
\dd s^2_{\text{AdS}_5\times\text{S}^5} &=& R^2 \bigg(\frac{-\dd t^2 +  \dd z^2 + \dd x^i \dd x_i}{z^2} + \dd \Omega_5^2 \bigg)\, ,
\end{eqnarray}
where we defined $R^2 \equiv \sqrt{4 \pi  g_s N} \alpha'$, i.e. the common AdS$_5$ and S$^5$ radius.
The second, and final, step consists in taking the Carroll limit. Based on the Carroll contraction of the AdS algebra \cite{Bergshoeff:2015wma, Figueroa-OFarrill:2018ilb, Figueroa-OFarrill:2019sex}, this amount to rescaling the coordinates and the radius as motivated in the Supplemental Material \ref{app:motivating_Carroll}, namely 
\begin{eqnarray}
\label{coords_resc_1NH_2Carroll}
    z\to \frac{z}{c} \, , \qquad
    x^i \to \frac{x^i}{c} \, , \qquad
    R \to \frac{R}{c} \, . 
\end{eqnarray}
By plugging the rescaling \eqref{coords_resc_1NH_2Carroll} into the metric \eqref{AdS5xS5_metric}, and by taking $c\to 0$, we obtain the Carroll AdS$_5\times$S$^5$ geometry,
\begin{eqnarray}
\label{Carroll_NC_AdS5xS5_metric}
    \dd s^2_{\text{Carroll AdS}_5\times\text{S}^5} &=& \left(\frac{1}{c^2} h_{\mu\nu} + \tau_{\mu\nu}\right) \dd X^{\mu} \dd X^{\nu}\,  , \\
    \notag
    h_{\mu\nu} \dd X^{\mu} \dd X^{\nu} &=& \frac{R^2}{z^2} \left( \dd z^2 + \dd x^i \dd x_i \right) + R^2  \dd \Omega_5^2 \, , \\
    \notag
    \tau_{\mu\nu} \dd X^{\mu} \dd X^{\nu} &=& -\frac{R^2}{z^2} \dd t^2 \, , 
\end{eqnarray}
where $h_{\mu\nu}$ and $\tau_{\mu\nu}$ describe $\mathbb{H}^4 \times S^5$ and warped $\mathbb{R}$, respectively.

At this point, we need to repeat these steps with the order of limits inverted, and show that we get the same final result. First, we rescale the coordinates and the radius as in \eqref{coords_resc_1NH_2Carroll}, but in addition we also rescale $\alpha'$ , namely 
\begin{eqnarray}
\label{coords_resc_1Carroll_2NH}
    \alpha' \to \frac{\alpha'}{c^2}\, , \qquad
    z\to \frac{z}{c} \, , \qquad
    x^i \to \frac{x^i}{c} \, , \qquad
    R \to \frac{R}{c} \, . 
\end{eqnarray}
Here there is an order of limits to resolve. Similarly to the argument provided in \cite{Fontanella:2024rvn}, we demand that the ``old'' $\alpha'$ still goes to zero. This is guaranteed provided that the ``new'' $\alpha'$ goes to zero faster than $c^2$.

Another important point to stress, is that within this rescaling the quantity $g_s N$ remains finite. This can be seen directly from the definition of $R$. Because $g_s = e^{\Phi}$, and the dilaton $\Phi$ is expected to rescale in the same way both in the gravity and gauge pictures,  $g_s$ needs to rescale as in \eqref{dilaton_resc_gauge}. Since $g_s N$ needs to remain finite, $N$ has to scale as well. In summary, $g_s$ and $N$ will have to scale as 
\begin{eqnarray}\label{gs_N_gravity}
    g_s \to c g_s \, , \qquad
    N\to \frac{N}{c} \, .
\end{eqnarray}
i.e. the number of D3-branes $N$ will be large in the $c\to 0$ limit, whereas the string coupling $g_s$ will be small.  

At this point, we plug the rescaling \eqref{coords_resc_1Carroll_2NH} into the metric of a stack of $N$ D3-branes \eqref{D3_metric_z} and take the $c\to 0$ limit, obtaining a Carroll D3-brane metric, 
\begin{eqnarray}
\label{Carroll_NC_D3_metric}
    \dd s^2_{\text{Carroll D3-brane}} &=&  \left(\frac{1}{c^2} h_{\mu\nu} + \tau_{\mu\nu}\right) \dd X^{\mu} \dd X^{\nu}\, , \\
    \notag
    h_{\mu\nu} \dd X^{\mu} \dd X^{\nu} &=& \frac{R^4}{\alpha^{\prime 2}\sqrt{f(z)}} \dd x^i \dd x_i + \frac{\alpha^{\prime 2} \sqrt{f(z)}}{z^4} \dd z^2 \\
    \notag
    &+& \frac{\alpha^{\prime 2} \sqrt{f(z)}}{z^2}  \dd \Omega_5^2\, , \\
    \tau_{\mu\nu} \dd X^{\mu} \dd X^{\nu} &=& -\frac{R^4}{\alpha^{\prime 2}\sqrt{f(z)}} \dd t^2  \, . \notag
\end{eqnarray}
The final step is to take the near-horizon limit $\alpha'\to0$ of \eqref{Carroll_NC_D3_metric}, which gives us precisely the same final result obtained in \eqref{Carroll_NC_AdS5xS5_metric}, as required for consistency. Under these limits, the string world-sheet remains relativistic, and the theory in the gravity picture corresponds to the magnetic Carroll string action \cite{Cardona:2016ytk, Harksen:2024bnh}, with target space \eqref{Carroll_NC_AdS5xS5_metric}. 

The full AdS$_5\times$S$^5$ type IIB supergravity solution is supported by fluxes, in particular it has a non-vanishing self-dual 5-form RR field strength $F^{(5)}$, and a constant dilaton $\Phi$. We analyse here the effect of the Carroll limit on them. The 5-form RR field strength is 
\begin{align}\label{5_form_RR}
    F^{(5)} &= \frac{1}{g_s R} \left[ \text{dvol}(\text{AdS}_5) + \text{dvol}(\text{S}^5) \right]  \\
    \notag
    \text{dvol}(\text{AdS}_5) &=  \frac{R^5}{z^5} \dd t\wedge \dd z\wedge \dd x^1 \wedge\dd x^2 \wedge\dd x^3 \, ,  \\
    \notag
    \text{dvol}(\text{S}^5) &= R^5 \sqrt{\frac{1-\frac{y^2}{4}}{\left( 1+\frac{y^2}{4} \right)^9}} \dd \phi \wedge \dd y_1 \wedge \dd y_2\wedge \dd y_3\wedge \dd y_4  \, . \notag
\end{align} 
By plugging the rescalings \eqref{coords_resc_1Carroll_2NH} and \eqref{gs_N_gravity} into \eqref{5_form_RR}, we get that the 5-form RR field solution will be 
\begin{eqnarray}
   F^{(5)} &=& \frac{1}{c^4} U^{(5)} + \frac{1}{c^5} V^{(5)} \, , \\
   \notag 
   U^{(5)} &=&  \frac{1}{g_s R}  \text{dvol}(\text{AdS}_5) \, , \qquad
   V^{(5)} =  \frac{1}{g_s R}  \text{dvol}(\text{S}^5) \, .
\end{eqnarray}
At this stage, it would be useful to check that our limit solutions are actual solutions of the supergravity equations of motion, and compute the Killing vectors preserved by the full solution, as required by holography. To perform both tasks, one should know the \emph{Carrollian Type IIB supergravity}, something that has not been developed yet.

\paragraph*{The Penrose boundary} 

The Penrose formalism can be applied to Newton-Cartan manifolds as well. We immediately see that the Carroll AdS$_5\times$S$^5$ metric \eqref{Carroll_NC_AdS5xS5_metric} measures an infinite distance at the point $z=0$. Therefore, we can set $\Omega = z$, and define \footnote{Here $\tau_{\mu\nu}$ is a representative of an equivalence class of metric tensors defined up to Carroll boosts. The conformal rescaling \eqref{conformal_resc} is understood to act on all elements of the equivalence class. Moreover, a generic Carroll boost shifts the longitudinal vielbein $\tau_{\mu}$ with a term proportional to the transverse vielbein $e_{\mu}^a$. That means the conformal rescaling cannot be gauged away with a Carroll boost.}   
\begin{eqnarray}\label{conformal_resc}
    \tilde{h}_{\mu\nu} = \Omega^2 h_{\mu\nu} \, , \qquad
    \tilde{\tau}_{\mu\nu} = \Omega^2 \tau_{\mu\nu} \, .
\end{eqnarray}
The conformal boundary is located at $z=0$, and the conformally rescaled metric tensors evaluated at the boundary read
\begin{eqnarray}\label{Penrose_CB}
        \left. \tilde{h}_{\mu\nu} \dd X^{\mu} \dd X^{\nu}\right|_{z=0} &= R^2 \dd x^i \dd x_i  \, , \\
        \notag
   \left. \tilde{\tau}_{\mu\nu} \dd X^{\mu} \dd X^{\nu}\right|_{z=0} &= - R^2 \dd t^2 \, . 
\end{eqnarray}
From this, we learn that the Penrose conformal boundary of Carroll AdS$_5\times$S$^5$ is the Carroll geometry of 4d flat spacetime.


\nicesection{Carroll limit in the gauge theory side}

We shall now incorporate the Carroll limit into the ``gauge theory side'' of Maldacena's holographic construction. The principle that we will follow, again, is to demand that the low energy limit $\alpha' \to 0$ commutes with the Carroll limit $c\to 0$. This consistency requirement is crucial, as it will rule out some possible rescalings. 

The system in the regime $g_s N \ll 1$ is described by interacting open and closed strings. The closed string action is given by type IIB superstring, which at low energy $\alpha'\to 0$ is captured by supergravity. The open string and the interaction term is instead described by the non-abelian DBI action \cite{Tseytlin:1997csa,Myers:1999ps}, see also the Supplemental Material \ref{app:NA_DBI}.

The low energy limit consists in expanding the DBI action around the flat spacetime metric and constant dilaton. We are also free to turn on a constant Kalb-Ramond B-field. However, in the context of the Carroll limit, it does not play any role, because the leading term in the metric is $h_{\mu\nu}$, a $9\times 9$ symmetric tensor \footnote{This is opposite to the non-relativistic limit, where in both gauge and string theories the B-field is crucial to cancel the divergence arising from the $2\times 2$ $\tau_{\mu\nu}$ metric, gaining the physical meaning of removing the ``rest energy'' near BPS bounds. In the Carroll limit this picture ceases to hold.}.  
Therefore, the background around which we expand the DBI is:
\begin{eqnarray}
    g_{\mu\nu} = \eta_{\mu\nu} \, , \qquad
    \Phi = \text{const.} \, , \qquad
    B = 0 \, , 
\end{eqnarray}
Guided by the fact the Carroll gauge theory needs to be holographically defined on the Penrose conformal boundary \eqref{Penrose_CB} - i.e. a Carroll Mink$_4$ spacetime - implies that the Carroll limit on the DBI should make the time coordinate $X^0$ smaller than the spatial directions $(X^i, X^I)$. One possibility to implement such rescaling is 
\begin{eqnarray}\label{Carroll_DBI_1}
    X^0 \to c X^0 \, , \qquad
    X^i \to X^i \, , \qquad
    X^I \to X^I \, ,
\end{eqnarray}
($i=1,...,3$; $I=4,..., 9$) and then take $c\to 0$. However, this is not the only possibility. For instance, one could instead rescale the space coordinates, and make them bigger than the time coordinate, 
\begin{eqnarray}\label{Carroll_DBI_2}
    X^0 \to  X^0 \, , \qquad
    X^i \to \frac{X^i}{c}  \, , \qquad
    X^I \to \frac{X^I}{c}  \, .
\end{eqnarray}
However, the rescaling \eqref{Carroll_DBI_2} does not produce a consistent limit, because it does not commute with  $\alpha'\to 0$.  On the other hand, the rescaling \eqref{Carroll_DBI_1} produces a consistent limit, and therefore we shall discuss this choice. In addition to \eqref{Carroll_DBI_1}, one also needs to rescale $g_s$ as 
\begin{eqnarray}\label{dilaton_resc_gauge}
    g_s \to c g_s \, .
\end{eqnarray}
Once the rescaligs \eqref{Carroll_DBI_1} and \eqref{dilaton_resc_gauge} are applied to the low energy limit $\alpha'\to 0$ of the DBI action, one obtains the bosonic sector of the magnetic Carroll $\mathcal{N}=4$ SYM \footnote{The full magnetic Carroll $\mathcal{N}=4$ SYM action, including fermions, is still not known, due to the lack of understanding of magnetic Carroll supersymmetry. The action proposed in \eqref{Carroll_N=4_SYM} captures the magnetic Carroll limit of only the bosonic sector of $\mathcal{N}=4$ SYM.}:  
\begin{eqnarray}\label{Carroll_N=4_SYM}
\notag
    S_{\text{CSYM}} &=& \frac{1}{2\pi g_s}\int \dd t\, \dd^{3}x \left[-\frac{1}{4} (F^{ij\, a})^2 
    - \frac{1}{2} ({\cal D}_i S^{I\, a} )^2 \right. \\
\notag
    &-& \left. \frac{1}{4} \kappa^{i \, a} F^a_{t i} + \frac{1}{2} \lambda^a_I{\cal D}_t S^{I\, a}  + \frac{1}{4} \left( \mathfrak{f}_{bc}{}^a S^{I\, b} S^{J\, c} \right)^2 \right]\, . \\
\end{eqnarray}
Here $\mathcal{D}$ is the $U(N)$ covariant derivative; $S^{I\, a}$ are six $U(N)$ adjoint scalars;  $\kappa^{i \, a}$ and $\lambda^a_I$ are Lagrange multiplier fields that dynamically impose the constraints: 
\begin{eqnarray}\label{constraints}
  F^a_{t i} = 0 \, , \qquad\qquad
  {\cal D}_t S^{I\, a} = 0 \, . 
\end{eqnarray}
Implicitly, the action \eqref{Carroll_N=4_SYM} assumes $N$ to be infinite, since in the gravity picture $N$ scales as $1/c$ in \eqref{gs_N_gravity}, and the same behavior is expected in the gauge picture. We leave a discussion of this effect for future work.


\nicesection{Symmetries of Carroll $\mathcal{N}=4$ SYM}

A first test of holography is to match symmetries. Therefore, we want to fix all spacetime symmetries of the Carroll $\mathcal{N}=4$ SYM action \eqref{Carroll_N=4_SYM} \footnote{The symmetries of the Carroll Yang-Mills action have also been studied in \cite{Islam:2023rnc}, although the internal symmetries acting on the Lagrange multipliers have not been discussed.}. The full analysis of the symmetries is reported in the Supplemental Material, here we summarise the results. The spacetime symmetries are generated by:
\begin{subequations}
\begin{eqnarray}
   K_i &=& 2x_i t \partial_t - x_jx^j\partial_i+ 2x_i x^j\partial_j \, , \\
   D&=&t\partial_t + x^i\partial_i \, , \\
  P_i&=&\partial_i \, , \\
   L_{ij}&=& x_j\partial_i-x_i\partial_j \, , \\
    M^{(l,m,n)}&=&x^ly^mz^n \partial_t \, .
\end{eqnarray}
\end{subequations}
where $x_1 \equiv x, x_2 \equiv y, x_3 \equiv z$.
These are precisely the generators of the Conformal Carroll Algebra (CCA) defined through \eqref{eq:CCAbrackets}. There are also purely internal symmetries which act on the fields exclusively. 
\begin{subequations}\label{eq:CYMinternal}
\begin{align}
\mathcal{I}&=\epsilon^{ijk} F_{jk} \frac{\partial}{\partial \kappa^i} \label{eq:outlier} \, ,\\
 N_{(ik)j}^{(l,m,n,p)} &=t^p\bigg[\partial_j(x^ly^mz^n)\; F_{tk}  + 2 x^ly^mz^n\; \partial_jF_{tk}\bigg]\frac{\partial}{\partial \kappa^i} \, ,\\
Q_{(ij)}^{(l,m,n,p)}&=x^ly^mz^n\left(pt^{p-1}\; F_{tj} + 2t^p\; \partial_t F_{tj}\right)\frac{\partial}{\partial \kappa^i}  \, ,\\
  R_{ij}^{(l,m,n,p)} &=t^px^ly^mz^n\left(F_{ti} \frac{\partial}{\partial \kappa^j} - F_{tj} \frac{\partial}{\partial \kappa^i}\right) \, ,\\
T_{[ij]I}^{(l,m,n)}&= 4\mathcal{D}_i\left(x^ly^mz^nS_I\right)\frac{\partial}{\partial \kappa^j} + x^ly^mz^nF_{ij} \frac{\partial}{\partial \lambda^I} \, ,\\
       U_{iI}^{(l,m,n,p)}&= t^p x^ly^mz^n\left(2\mathcal{D}_t S_I\frac{\partial}{\partial \kappa^i}+F_{ti} \frac{\partial}{\partial \lambda^I}\right) \, ,\\
        V_{[IJ]}^{(l,m,n)} &= \bigg[x^ly^mz^n\lambda^J+2t\partial_i\partial^i (x^ly^mz^n) S_J \nonumber \\
        &\qquad \qquad \;\;\;+4t\partial^i (x^ly^mz^n) \mathcal{D}_iS_J\bigg]\frac{\partial}{\partial \lambda^I}  \\
        & \hspace{-10mm}+ 4t \partial^i (x^ly^mz^n)\left[S_I, S_J\right]\frac{\partial}{\partial \kappa^i}  +x^ly^mz^nS^J\frac{\partial}{\partial S^I} \, ,\nonumber
        \label{eq:Vgenerator} 
        \end{align}
         \begin{align}
    W_{i[IJ]}^{(l,m,n)}&= \bigg[\partial_i (x^ly^mz^n)S_J+ 2x^ly^mz^n\mathcal{D}_iS_J\bigg] \frac{\partial}{\partial \lambda^I}\nonumber \\
         &\qquad \;+ 2x^ly^mz^n\left[S_I, S_J\right]\frac{\partial}{\partial \kappa^i} \, ,\\
   X_{IJ}^{(l,m,n)}&=x^ly^mz^n\bigg[S_J \frac{\partial}{\partial \lambda^I}+S_I \frac{\partial}{\partial \lambda^J}\bigg]  \, ,\\
        Y_{(IJ)}^{(l,m,n,p)}&=x^ly^mz^n\bigg[\partial_t (t^p\mathcal{D}_t S_J)+t^p \partial_t \mathcal{D}_t S_J\bigg]\frac{\partial}{\partial \lambda^I} \\
       Z_{i(IJ)}^{(l,m,n,p)}&=t^p\bigg[\partial_i (x^ly^mz^n\mathcal{D}_t S_J) + x^ly^mz^n\partial_i \mathcal{D}_t S_J\bigg]\frac{\partial}{\partial \lambda^I} \, ,\\
       J_{IJ}^{(l,m,n,p)} &=t^p x^ly^mz^n\bigg[\mathcal{D}_t S_J \frac{\partial}{\partial \lambda^I}-\mathcal{D}_t S_I \frac{\partial}{\partial \lambda^J}\bigg]\, .
    \end{align}
\end{subequations}
The notation used is the following. $K_{(ij)k}$ or $K_{[ij]k}$ means that the generator is symmetric (resp. anti-symmetric) in the $i, j$ indices, even though not explicitly spelled-out. Moreover, any generator of the form $A \frac{\partial}{\partial B}$ should be implemented in the action in terms of infinitesimal transformations as $\delta B = \alpha A$, with $\alpha$ an infinitesimal parameter. Generators that carry an index structure, e.g. $A_{(I} \frac{\partial}{\partial B^{J)}}$, translates as $\delta B^I = \omega_{IJ} A^J$, where $\omega_{IJ}$ is a symmetric rank-2 tensor of infinitesimal parameters. For convenience, in the above expressions the gauge indices have been suppressed in the form $S_I = S_I^a T_a$, with $T_a$ being the $U(N)$ generators.

It is interesting to note that a subset of the internal symmetries, precisely the one given in \eqref{eq:badguys}, does not close into another symmetry under ordinary Lie brackets. The commutator between the internal and CCA generators produce, in general, new generators that are not symmetries of the action, see e.g. \eqref{eq:nonclosure} and \eqref{eq:nonclosure_2}. This might have to do with the fact that the Lagrange multipliers transform non-linearly, although we do not have a full understanding of why this is happening.   

It might be possible that the Noether charges associated with many of these symmetries are vanishing, similarly to what found in the cases discussed in \cite{Lambert:2024uue, Lambert:2024yjk}. When the Noether charge associated to a certain symmetry vanishes, then such symmetry is regarded as `unphysical', or as a gauge redundancy, and it is not expected to be holographically matched in the dual gravity theory. Computing the Noether charges associated with the above symmetries is beyond the scope of this letter, and it is left for future work.

In addition to these internal symmetries, there are also the usual $U(N)$ gauge symmetries, which are preserved as the Carroll limit did not act on gauge indices.


\nicesection{Conclusions}
\label{sec:conclusions}

In this letter we construct a novel holographic duality involving a Carroll bulk theory. The duality is derived via incorporating the Carroll limit inside the AdS/CFT setting of Maldacena and by demanding consistency, namely the gauge theory is defined on the Penrose boundary of the bulk theory and the Maldacena's decoupling limit commutes with the Carroll limit. 

Due to the current limitation of knowledge, we are unable to check that the symmetries of the Carroll $\mathcal{N}=4$ SYM, which we presented in this letter, match with the symmetries of the bulk theory. The symmetries of the bulk theory are given by the Killing vectors that leave the full solution of Carroll type IIB supergravity invariant, up to local symmetries. It is of paramount importance to develop Carroll type IIB supergavity in order to check that the proposed limit solution solves the equations of motion and to find its symmetries \footnote{At present, only non-relativistic type IIB supergravity has been worked out \cite{Bergshoeff:2023ogz}. This has been instrumental in the symmetry match proof of the non-relativistic AdS/CFT correspondence \cite{Fontanella:2024kyl, Fontanella:2024rvn, Fontanella:2024hgv, Lambert:2024uue, Lambert:2024yjk, Lambert:2024ncn}. Although not in the context of the type IIB theory, some progress in Carroll supergravity has been done in \cite{Grumiller:2024dql, Ravera:2019ize, Ravera:2022buz, Ali:2019jjp}}.

Our proposed Carroll bulk holography creates the need of a more rigorous formulation of Carroll string theory. In recent years, there has been some progress in this subject, see e.g. \cite{Cardona:2016ytk, Harksen:2024bnh, Casalbuoni:2024jmj, Bagchi:2023cfp, Bagchi:2024rje, Bagchi:2024qsb, Bagchi:2025vri, Banerjee:2023ekd, Banerjee:2024fbi, Blair:2023noj, Gomis:2023eav, Blair:2025nno}. However, there are some fundamental questions that are still unanswered. For example, the Carroll limit in target space produces a divergent term in the Polyakov action that is tamed via a Hubbard-Stratonovich transformation. However, in the Nambu-Goto action  there is no B-field that can cancel the divergent term, and the Hubbard-Stratonovich transformation is of unclear application, due to the square root. It is a fundamental open question to show how these two relativistically equivalent string actions are still equivalent after taking the Carroll limit \footnote{For a recent progress in this direction, see \cite{Gomis:2023eav}.}.

An interesting quantitative test of the proposed duality would be to match the Carroll string spectrum with the scaling dimensions of gauge invariant operators of the Carroll $\mathcal{N}=4$ SYM. In the case of the non-relativistic limit, the theory is zooming into a BPS bound, featuring miraculous cancellations in the string spectrum, leading to 1-loop exact formulas \cite{Fontanella:2021btt, Fontanella:2021hcb, Fontanella:2023men, deLeeuw:2024uaq}. We do not expect the same to happen for the Carroll string, as it is not associated with zooming into BPS bounds.

Another important direction is to add fermions to our description. Carroll fermions have been studied in \cite{Bergshoeff:2023vfd, Banerjee:2022ocj, Bagchi:2022eui, Koutrolikos:2023evq}. However, it is still an open problem how to couple them supersymmetrically to a bosonic action when the Carroll limit is of magnetic type. Our proposed holography gives a ground motivation to explore this issue, since the gauge theory appears to be in the \emph{magnetic} Carroll limit. The main issue with magnetic Carroll supersymmetry is that the commutator of two supersymmetry transformations does not close into a translation. It would be interesting to explore if this is connected to the non-closure of the symmetry algebra found in this letter. Perhaps one needs to introduce new modified brackets that take into account the constraints produced by the Lagrange multipliers, similarly to the Dirac brackets in the Hamiltonian quantisation. The electric Carroll limit does not involve Lagrange multipliers, suggesting why the issue of non-closure does not occur there.     

Regarding the symmetries of the Carroll $\mathcal{N}=4$ SYM action, it would be important to study their Noether charges. It might be possible that many of the symmetries found in this letter have vanishing charge, and therefore can be discarded for being 'unphysical'. This would have a holographic consequence, as only the physical charges are expected to be matched in the dual gravity theory. We leave this investigation for future work.  

The proposed triality between Carroll string theory, relativistic strings in flat spacetime and Carroll $\mathcal{N}=4$ SYM, potentially, provides a new window to look into the problem of flat space holography \footnote{See also \cite{Banerjee:2025tec} for an approach to derive celestial CFTs from the AdS/CFT, and \cite{Filippas:2024tgs} for a nesting of flat space into the AdS/CFT.}. This map provides a concrete setting where infrared-divergent observables in flat space can be computed in Carroll-AdS, where such infrared problems are naturally tamed, e.g. the 1-loop partition function, as commented in \cite{Matulich:2019cdo}.     
This naturally raises the question of what is the duality that relates Carroll string theory and relativistic string theory in the compactified flat spacetime. For non-relativistic string theory, T-duality maps a non-relativistic into a relativistic action \cite{Gomis:2005pg, Bergshoeff:2018yvt}. This is possible due to the presence of the critical B-field playing a crucial role in the Buscher rules. For Carroll string theory, this is no longer the case, and therefore this issue deserves a more refined idea. An interesting direction to explore is given by the fact Carroll AdS is isomorphic to the blow up of time-like infinity in flat spacetime, $\mathsf{Ti}$ \cite{Figueroa-OFarrill:2021sxz}. It would be interesting to understand how the internal space $X^5$ behaves under the blow up. By requiring to match Carroll AdS$_5\times$S$^5$, together with matching the physical symmetries of Carroll $\mathcal{N}=4$ SYM,  we expect this will fix the $X^5$ geometry.

Finally, it would be interesting to see if the Carroll bulk holography proposed in this letter has any application in the context of a hypothetical de Sitter holography, as suggested in \cite{Blair:2025nno, Argandona:2025jhg}.

We hope that the results presented in this letter offer a new perspective on the problem of flat space holography, and provide strong motivation to initiate a systematic study of Carroll string theory and related topics.


{\bf Acknowledgements:} We would like to thank J.M. Nieto Garc\'ia  for useful discussions and for feedback on a draft of this letter. AF thanks J. M. Nieto Garc\'ia for an initial collaboration on this project. 
AF is supported by the SFI and the Royal Society under the grant number RFF$\backslash$EREF$\backslash$210373.
AF thanks Lia for her permanent support.


\bibliographystyle{apsrev4-1}

\bibliography{Biblio}

\onecolumngrid
\clearpage
\begin{center}
\textbf{\large Supplemental Material}
\end{center}
\vspace{0.2cm}
\setcounter{section}{0}
\renewcommand{\thesection}{S\arabic{section}}
\setcounter{equation}{0}
\renewcommand{\theequation}{S\arabic{equation}}
\setcounter{figure}{0}
\renewcommand{\thefigure}{S\arabic{figure}}
\setcounter{table}{0}
\renewcommand{\thetable}{S\arabic{table}}

\section{Non-Abelian DBI and Notations}
\label{app:NA_DBI}

In the weak string coupling regime of the Maldacena's setting, the system is described by interacting closed and open strings. The open string and the interaction terms are described by the non-abelian DBI action \cite{Tseytlin:1997csa,Myers:1999ps},
\begin{equation}\label{DBI}
    S_{\text{DBI}}=-T_4 \int{\dd x^4 \Tr_S \left[ e^{-\Phi} \sqrt{-\det (Q^I{}_J) \det \left( \Pu_c\left[ E_{\mu\nu} + E_{\mu I} (Q^{-1}-\delta)^I\null_\nu  \right]_{\alpha\beta} + 2\pi \alpha' F_{\alpha\beta}  \right)} \right] } \, , 
\end{equation}
where $T_4$ is the D3-brane tension, $E\equiv g+B$ is the combination of the metric and the Kalb-Ramond B-field; $x^{\alpha}$, with $\alpha = 0,..., 3$, are the brane world-volume coordinates; the indices $\mu, \nu, ...$ label coordinates of the 10d target space; $F_{\alpha\beta} = \partial_{\alpha} A_{\beta} -\partial_{\beta} A_{\alpha} +i [A_{\alpha} , A_{\beta}]$ is the field strength of a gauge potential $A_{\alpha} = A_{\alpha}^a T^a$, where $T^a$ are generators of $U(N)$ that carry the Chan-Paton indices; $\Pu_c [E_{\mu\nu}]_{\alpha\beta}$ indicates the pullback of $E_{\mu\nu}$, with the usual derivatives replaced by covariant derivatives $D_{\alpha} X^{\mu}= \partial_{\alpha} X^{\mu} + i [A_{\alpha} , X^{\mu}]$; and $\Tr_S$ is the symmetrised trace over Chan-Paton indices, that is, the trace over the sum of all the ways we can permute the matrices
\begin{equation}
    \Tr_S (A_1 \dots A_n) = \Tr( A_1 \dots A_n + \text{perm.}) \, .
\end{equation}
Finally, the matrix $Q$ takes the form
\begin{equation}
    Q^I\null_J = \delta^I\null_J +\frac{i [X^I , X^K]}{2\pi \alpha '} E_{KJ} \, ,
\end{equation}
where the indices $I$, $J$ and $K$ run over the spacetime coordinates that are transverse to the D3-brane, i.e. $I,J,K = 4, ..., 9$. We denote the spacetime coordinates longitudinal to the D3-brane as $(X^0, X^i)$, with $i=1,2,3$.

The decoupling limit consists in zooming near the D3-brane world-volume, which is achived by rescaling the transverse fields as 
\begin{eqnarray}
    X^I = 2\pi \alpha' S^I \, , 
\end{eqnarray}
and by taking $\alpha'\to 0$. Then, the $S^I$ are the transverse fields entering in the $\mathcal{N}=4$ SYM action.

\section{Motivation of the Carroll limit taken in the bulk}
\label{app:motivating_Carroll}

The starting point is the AdS$_5\times$S$^5$ geometry, which has isometry $\mathfrak{so}(2,4) \oplus \mathfrak{so}(6)$. This algebra is given by the commutation relations, 
\begin{subequations}\label{so(4,2)+so(6)}
	\begin{align}
    \notag
    &\text{AdS}_5  & &  \text{S}^5 \\
	[P_{\hat{a}}, P_{\hat{b}}] &= \frac{1}{R^2} J_{\hat{a}\hat{b}} \ , &
	[P_{a'}, P_{b'}] &= - \frac{1}{R^2} J_{a'b'} \ , \\
	[P_{\hat{a}}, J_{\hat{b}\hat{c}}] &= 2 \eta_{\hat{a}[\hat{b}} P_{\hat{c}]} \ , &
	[P_{a'}, J_{b'c'}] &= 2 \delta_{a'[b'} P_{c']} \ , \\
	[J_{\hat{a}\hat{b}}, J_{\hat{c}\hat{d}}] &= 2\eta_{\hat{c}[\hat{b}} J_{\hat{a}]\hat{d}} - 2\eta_{\hat{d}[\hat{b}} J_{\hat{a}]\hat{c}}\ , &
	[J_{a'b'}, J_{c'd'}] &= 2\delta_{c'[b'} J_{a']d'} 
    - 2\delta_{d'[b'} J_{a']c']}\ ,
	\end{align}
\end{subequations}
where $\hat{a}, \hat{b}, ...$ are indices of $\mathfrak{so}(2,4)$ that run from 0 to 4, whereas $a', b', ....$ are indices of $\mathfrak{so}(6)$ that run from 5 to 9. 

As pointed out in \cite{Cardona:2016ytk}, the \emph{stringy} and the \emph{particle} Carroll limits yields to the same physics, in sharp contrast to the case of the non-relativistic limit. 
From the algebraic point of view, this translates into taking a particle Carroll contraction of the AdS$_5\times$S$^5$ isometry algebras. 
The particle Carroll contraction of the AdS algebra was discussed in \cite{Bergshoeff:2015wma}. We generalise it to AdS$_5\times$S$^5$ by treating the sphere directions $a'$ on the same footing as the transverse (spatial) direction of AdS, implementing the idea that time scales different from space.  Concretely, this amount to splitting any AdS$_5$ index as $\hat{a} = (0, a)$, with $a = 1, ..., 4$ being the transverse index, and by rescaling the generators as 
\begin{eqnarray}
\label{rescaling_1}
    P_0 = \frac{1}{c} H \, , \qquad
    J_{0 a} = \frac{1}{c} G_a \, , 
\end{eqnarray}
whereas we do not rescale any generator of the S$^5$ algebra, since there are no generators of the type $J_{0a'}$.  
Finally, we take $c\to 0$, and obtain the AdS$_5\times$S$^5$ Carroll algebra: 
\begin{subequations}\label{Carroll_AdSxS}
	\begin{align}
    \notag
    &\text{AdS}_5  & &  \text{S}^5 \\
	[P_a, P_b] &= \frac{1}{R^2} J_{ab} \ , &
	[P_{a'}, P_{b'}] &= - \frac{1}{R^2} J_{a'b'} \ , \\
	[P_a, J_{bc}] &= 2 \delta_{a[b} P_{c]} \ , &
	[P_{a'}, J_{b'c'}] &= 2 \delta_{a'[b'} P_{c']} \ , \\
	[J_{ab}, J_{cd}] &= 2\delta_{c[b} J_{a]d} - 2\delta_{d[b} J_{a]c}\ , &
	[J_{a'b'}, J_{c'd'}] &= 2\delta_{c'[b'} J_{a']d'} 
    - 2\delta_{d'[b'} J_{a']c']}\ . \\
    [H, P_a] &= \frac{1}{R^2} G_a  \ , &
	&  \\
    [P_a, G_b] &= - \delta_{ab} H \ , &
	&  \\
    [G_a, J_{bc}] &= \delta_{ab} G_c - \delta_{ac} G_b  \ , &
	&  
	\end{align}
\end{subequations}
There is a second ``complementary'' rescaling of generators that leads to the same algebra \eqref{Carroll_AdSxS} in the $c\to 0$ limit. This is given by 
\begin{eqnarray}
\label{rescaling_2}
    P_a = c \tilde{P}_a \, , \qquad
    P_{a'} = c \tilde{P}_{a'} \, , \qquad
    J_{0 a} = \frac{1}{c} G_a \, , \qquad
    R = \frac{\tilde{R}}{c} \, . 
\end{eqnarray}
This second possibility will be useful in a moment. Now that we have identified the contraction of the isometry algebra, we move to the metric. The goal is to identify the coordinate rescaling associated to the above algebra contraction. 

We start with the AdS$_5\times$S$^5$ metric given in \eqref{AdS5xS5_metric}. The rescaling of coordinates induced by \eqref{rescaling_1} is
\begin{eqnarray}
\label{coords_resc_1}
  t \to c t \, .  
\end{eqnarray}
However, once this rescaling is plugged into the Polyakov string action, it is immediate to see that the time field $t$ will disappear from the action in the $c\to 0$ limit. 

The alternative coordinate rescaling induced by \eqref{rescaling_2} is
\begin{eqnarray}
\label{coords_resc_2}
    z\to \frac{z}{c} \, , \qquad
    x^i \to \frac{x^i}{c} \, , \qquad
    R \to \frac{R}{c} \, . 
\end{eqnarray}
where the S$^5$ coordinates do not need to be rescaled, as they are dimensionless. This second rescaling keeps all fields in the Polyakov action, creating a divergent term in the $c\to 0$ limit which is tamed by the Hubbard-Stratonovich transformation. Since we want to retain all fields in our action, we choose the rescaling \eqref{coords_resc_2}.

\newpage
\section{Symmetries of Carroll $\mathcal{N}=4$ SYM}
\label{sec:CSYMsymm}

The bosonic sector of the Carroll $\mathcal{N}=4$ Super-Yang Mills action \eqref{Carroll_N=4_SYM} was found as a `magnetic' Carroll limit of the relativistic $\mathcal{N}=4$ Super-Yang Mills. To deal with the divergent terms, we applied the Hubbard-Stratonovich transformation, which introduced the Lagrange multiplier fields $\kappa^{i \, a}$ and $\lambda_I^a$.

To find the symmetries of this action, we first write down the most general infinitesimal field transformations in \eqref{eq:fieldtransformations}. Associated to each transformation there are the parameters in Greek script which depend on the spacetime coordinates, with the spacetime coordinates transforming as $x^\mu \rightarrow x^\mu - \xi^\mu$, i.e. $\delta x^\mu = -\xi^\mu$. One particularly interesting point is that because $\kappa^{i \, a}$ and $\lambda_I^a$ act as Lagrange multipliers (they are not dynamical fields), one must go beyond linear order in the fields and derivatives for these transformations, to capture all the spacetime and internal symmetries. This was also noted in \cite{Islam:2023rnc}, although below we also report the internal symmetries of this theory.

For notational convenience, we have written a bar $-$ to indicate the $t$-component, while no bar indicates a spatial component
\begin{subequations}\label{eq:fieldtransformations}
\begin{align}
    \delta A_t = \;& \xi^t \partial_t A_t + \xi^j \partial_j A_t + \alpha_{\bar{A}} A_t + \alpha^j_{A} A_j + \alpha^j_{\kappa} \kappa_j + \alpha^{J}_{S}S_J + \alpha^J_{\lambda}\lambda_J \, ,   \\
    \delta A^i =\;& \xi^t \partial_t A^i + \xi^j \partial_j A^i + \beta^i_{\ba} A_t + \beta^{ij}_A A_j + \beta^{ij}_{\kappa} \kappa_j + \beta^{iJ}_{S}S_J + \beta^{iJ}_{\lambda}\lambda_J \, ,\\
    \delta S^I=\;&  \xi^t \partial_t S^I + \xi^j \partial_j S^I +  \delta^I_{\ba} A_t +  \delta^{jI}_AA_j + \delta^{jI}_\kappa\kappa_j + \delta^{IJ}_SS_J + \delta^{IJ}_\lambda\lambda_J \, ,\label{eq:Scalartransfo} \\
    \delta \kappa^i =\;& \xi^t \partial_t \kappa^i + \xi^j \partial_j \kappa^i + \gamma^i_{\ba} A_t + \gamma^{ij}_A A_j +\gamma^{ij}_\kappa \kappa_j + \gamma^{iJ}_SS_J + \gamma^{iJ}_S\lambda_J  +\gamma^i_{\bp\ba} \partial_t A_t + \gamma^{ij}_{\partial \ba} \partial_j A_t + \gamma^{ij}_{\bp A}\partial_t A_j + \gamma^{ijk}_{\partial A} \partial_j A_k\nonumber \\
     &+ \gamma^{iJ}_{\bp S}\partial_t S_J + \gamma^{ijJ}_{\partial S}\partial_j S_J + \gamma^{iJ}_{\bp \lambda}\partial_t \lambda_J + \gamma^{ijJ}_{\partial \lambda}\partial_j \lambda_J  + \gamma^i_{\bp \bp A} \partial_t \partial_t
    A_t + \gamma^{ij}_{\bp \partial \ba}\partial_t\partial_j A_t + \gamma^{ijk}_{\partial \partial \ba}\partial_j \partial_k A_t + \gamma^{ij}_{\bp\bp A}\partial_t \partial_t A_j \nonumber \\
    &+ \gamma^{ijk}_{\bp\partial A}\partial_t \partial_j A_k + \gamma^{ijkl}_{\partial \partial A}\partial_j \partial_k A_l+\gamma^{ij}_{\kappa\ba}[\kappa_j,A_t] + \gamma^{ijk}_{\kappa A}[\kappa_j, A_k]+\gamma^{ij}_{\ba A} [A_t, A_j] + \gamma^{ijk}_{AA}[A_j, A_k] +\gamma^{iJK}_{\lambda S} [\lambda_J, S_K] \nonumber \\
    &+ \gamma^{iJK}_{SS}[S_J, S_K] + \gamma^{iJ}_{\ba S} [A_t, S_J] + \gamma^{ijJ}_{AS}[A_j, S_J] + \gamma^{iJ}_{\ba\lambda} [A_t, \lambda_J] + \gamma^{ijJ}_{A\lambda}[A_j, \lambda^J] + \gamma^{ijJ}_{\kappa\lambda}[\kappa_j, \lambda_J]+ \gamma^{ij}_{\bp \ba A}\partial_t[A_t, A_j]  \nonumber \\
    &+ \gamma^{ijk}_{\partial \ba A}\partial_j[A_t, A_k] + \gamma^{ijk}_{\bp AA}\partial_t [A_j, A_k] + \gamma^{ijkl}_{\partial A A}\partial_j [A_k, A_l] + \gamma^{iJK}_{\bp SS} \partial_t [S_J, S_K] + \gamma^{ijJK}_{\partial SS}\partial_j [S_J, S_K]  \\
   &+\gamma^{iJK}_{\ba SS}[[A_t, S_J], S_K] + \gamma^{iJK}_{SS \ba}[[S_J, S_K], A_t] + \gamma^{ijJK}_{ASS}[[A_j, S_J], S_K] + \gamma^{ijJK}_{SSA}[[S_J, S_K], A_j] \, ,\nonumber \\
    \delta \lambda^I = \; &\xi^t \partial \lambda^I + \xi^j \partial_j \lambda^I + \zeta^I_{\ba} A_t + \zeta^{jI}_AA_j + \zeta^{jI}_\kappa\kappa_j + \zeta^{IJ}_SS_J + \zeta^{IJ}_\lambda\lambda_J +\zeta^I_{\bp\ba} \partial_t A_t + \zeta_{\partial \ba}^{jI} \partial_j A_t + \zeta_{\bp A}^{jI}\partial_t A_j + \zeta_{\partial A}^{jkI}\partial_j A_k \nonumber \\
    &+ \zeta_{\bp \kappa}^{jI}\partial_t \kappa_j + \zeta_{\partial \kappa}^{jkI}\partial_j \kappa_k +\zeta^{IJ}_{\bp S}\partial_t S_J + \zeta_{\partial S}^{jIJ}\partial_j S_J+\zeta_{\bp \lambda}^{IJ}\partial_t \lambda_J + \zeta_{\partial \lambda}^{jIJ}\partial_j \lambda_J + \zeta_{\bp \bp S}^{IJ}\partial_t \partial_t S_J + \zeta_{\bp \partial S}^{jIJ}\partial_t \partial_j S_J \nonumber\\
    &+ \zeta_{\partial \partial S}^{jkIJ}\partial_j \partial_k S_J+\zeta_{\kappa\ba}^{jI}[\kappa_j,A_t] + \zeta_{\kappa A}^{jkI}[\kappa_j, A_k]+\zeta_{\ba A}^{jI} [A_t, A_j] + \zeta_{AA}^{jkI}[A_j, A_k]+\zeta_{\ba S}^{IJ}[A_t, S_J]+\zeta_{A S}^{jIJ}[A_j, S_J] \nonumber \\
    & +\zeta_{\ba \lambda}[A_t, \lambda_J]+\zeta_{A \lambda}^{jIJ}[A_j, \lambda_J] +\zeta_{\kappa S}^{jIJ}[\kappa_j, S_J]+\zeta_{SS}^{IJK}[S_J, S_K]+\zeta_{S\lambda}^{IJK}[S_J, \lambda_K]+ \zeta_{\bp\ba S}^{IJ}\partial_t [A_t, S_J] \nonumber\\
    &+ \zeta_{\partial \ba S}^{jIJ}\partial_j [A_t, S_J]+\zeta_{\bp A S}^{jIJ}\partial_t [A_j, S_J]+ \zeta_{\partial A S}^{jkIJ}\partial_j [A_k, S_J] +\zeta_{SSS}^{IJKL} [[S_J,S_K], S_L] +\zeta_{\ba S \ba}^{IJ} [[A_t,S_J],A_t] \\
    &+ \zeta_{AS\ba}^{jIJ} [[A_j,S_J],A_t]+ \zeta_{\ba S A}^{jIJ} [[A_t,S_J],A_j] + \zeta_{\ba A S}^{jIJ} [[A_t,A_j],S_J]  + \zeta_{ASA}^{jkIJ} [[A_j,S_J],A_k]+ \zeta_{AAS}^{jkIJ} [[A_j,A_k],S_J] \, .\nonumber
\end{align}    
\end{subequations}
For these transformations to be symmetries of the theory, we require the Lagrangian to transform as a total derivative, i.e. $\delta \mathcal{L} = \partial_\mu F^\mu$ with $F^\mu$ being some functions of fields, coordinates and parameters. First, this fixes the vector field components $\xi^t$ and $\xi^i$ as
\begin{subequations}\label{eq:xi}
\begin{align}
    \xi^t &= (2B_i x^i+C) t + f(x) \, ,\\
    \xi^i &= D^i + Cx^i+ E^i{}_j x^j - B^ix^jx_j +2 B_j x^ix^j \, ,
\end{align}
\end{subequations}
with all the letters in capital Latin script constants and $f(x)$ an arbitrary function of the spatial coordinates only. Here also $E_{ij} = E_{[ij]}$. The other parameters in Greek script are now further constrained by derivatives of $\xi^\mu$ and the introduction of a few arbitrary functions of spacetime:
\begin{subequations}
\begin{align}
    \alpha_{\ba} &= C +2B_i x^i \, , \\
    \beta^i_{\ba}&= 2B^i t + \partial^i f(x) \, ,\\
    \beta^{ij}_A&= (C+2B_kx^k)\delta^{ij} - E^{ij}+2B^ix^j-2B^{j}x^i \, ,\\
    \gamma^{ij}_\kappa &=2(C+2B_kx^k)\delta^{ij} - E^{ij}+2B^ix^j-2B^{j}x^i \, , \\
    \gamma^{ij}_{\partial \ba} = -\gamma_{\bp A}^{ij}=  - \gamma_{\ba A}^{ij} &= k^{ij}(t,x)- \frac{1}{2}\bigg(\partial_k g^{ikj}(t,x) + \partial_t h^{ij}(t,x)\bigg) \, ,\\
    \gamma_{\partial A}^{ijk} = 2 \gamma_{AA}^{ijk}&= 4(2B^k t+ \partial^kf(x))\delta^{ij}- 4(2B^j t+ \partial^jf(x))\delta^{ik} - 2\epsilon^{ijk}G \, ,\\
    \gamma_{\bp \ba A}^{ij} = -  \gamma_{\bp\partial \ba}^{ij}= \gamma_{\bp\bp A}^{ij} &\equiv h^{ij}(t,x) \, ,\\
    \gamma_{\partial \ba A}^{ijk}= \gamma_{\bp\partial A}^{ijk} &\equiv g^{ijk}(t,x) \, ,\\
    \gamma_{\partial \partial \ba}^{ijk} &=  -\frac{1}{2}\bigg(g^{ikj}(t,x)+g^{ijk}(t,x) \bigg) \, ,\\
    \gamma_{S}^{iI} &= 4\partial_ja^{jiI}(x) \, ,\\
    \gamma_{\bp S}^{iJ} =\gamma_{\ba S}^{iJ}=-2\zeta_{\partial \ba}^{iJ}=2\zeta_{\bp A}^{iJ}=2\zeta_{\ba A}^{iJ}&\equiv2b^{iJ}(t,x) \, ,\\
    \gamma_{\partial S}^{ijI} =\gamma_{A S}^{ijI}=-2\zeta_{\partial A}^{ijJ}=-4\zeta_{AA}^{ijI} &\equiv -4a^{ijI}(x)\, ,\\
    \gamma_{SS}^{iIJ} &=4t\partial^ic^{IJ}(x)+ d^{iIJ}(x) \, ,\\
    \delta_S^{IJ}&= (2B_i x^i + C)\delta^{IJ}+ c^{IJ}(x)\, ,\\
    \zeta_S^{IJ} &= n^{IJ}(x) +\frac{1}{2}\partial_j d^{jIJ}(x)+2t\partial^j\partial_j c^{IJ}(x) \, ,\\
     \zeta_\lambda^{IJ} &= 2(2B_i x^i + C)\delta^{IJ} + c^{IJ}(x)\, ,\\
     \zeta_{\bp S}^{IJ}= \zeta_{\ba S}^{IJ} &= r^{IJ}(t,x)+ \frac{1}{2}\bigg( \partial_t p^{IJ}(t,x)+\partial_j q^{jIJ}(t,x) \bigg)\, ,\\
     \zeta_{\partial S}^{jIJ}=\zeta_{A S}^{jIJ} &=2(2B^j t + \partial^j f(x))\delta^{IJ} + 4t\partial^jc^{IJ}(x)+ d^{jIJ}(x) \, ,\\
    \zeta_{\bp \ba S}^{IJ} =\zeta_{\bp\bp S}^{IJ} &\equiv p^{IJ}(t,x)\, ,\\
     \zeta_{\partial \ba S}^{iIJ}=\zeta_{\bp \partial S}^{iIJ}&\equiv q^{iJK}(t,x) \, .
\end{align}
\end{subequations}
The new arbitrary functions are given by $a^{ijI} =  a^{[ij]I}$, $b^{iJ}$, $c^{IJ} = c^{[IJ]}$, $d^{iIJ} = d^{i[IJ]}$, $g^{ijk} = g^{(i|j|k)}$, $h^{ij} = h^{(ij)}$, $k^{ij}= k^{[ij]}$, $n^{IJ} = n^{(IJ)}$, $ p^{IJ}=p^{(IJ)}$, $q^{iIJ}=q^{i(IJ)}$, $r^{IJ}= r^{(IJ)}$. Their dependence on the spacetime coordinates are given in the expressions above. Additionally, $\delta^{ij}$ is the Kroenecker delta (not to be confused with the parameters of $\delta S^I $ in \eqref{eq:Scalartransfo}), $\epsilon^{ijk}$ is the Levi-Civita symbol with $\epsilon^{123} = 1$, and $G$ is a constant. Beyond this, we also have
\begin{equation}
    \gamma_{\ba S S}^{iJK} = 2\gamma_{SS\ba}^{iJK}, \qquad 
   \gamma_{ASS}^{ijJK} = 2\gamma_{SSA}^{ijJK}, \qquad 
     \zeta_{AS\ba}^{jIJ} = -\zeta_{\ba S A}^{jIJ}=\zeta_{\ba A S}^{jIJ}, \qquad 
     \zeta_{ASA}^{jkIJ} = 2 \zeta_{AAS}^{jkIJ}, 
\end{equation}
with $ \gamma_{\ba S S}^{iJK}= \gamma_{\ba S S}^{i[JK]}$, $  \gamma_{ASS}^{ijJK}=  \gamma_{ASS}^{ij[JK]}$, and $\zeta_{ASA}^{jkIJ} =  \zeta_{ASA}^{[jk]IJ}$. However these relations simply imply the Jacobi identity and hence these parameters and their corresponding terms in the transformation \eqref{eq:fieldtransformations} do not lead to new symmetries. All the remaining parameters are vanishing. We also investigated the shift symmetry, where each field is shifted by an arbitrary function of spacetime, e.g. $\delta A_t = f_{A_t}(t,x)$. However, each of these functions were constrained to be zero.
\newline

From these considerations, we can find the spacetime and internal symmetries of \eqref{Carroll_N=4_SYM}. The spacetime symmetries generate an infinite dimensional algebra spanned by the generators associated with the capital Latin (constant) transformation parameters $B^i,C,D^i, \dots$ in \eqref{eq:xi}. These can be found as
\begin{subequations}\label{eq:spacetimesymm}
\begin{alignat}{2}
    B^i:& \qquad \qquad \qquad \qquad  K_i &&= 2x_i t \partial_t - x_jx^j\partial_i+ 2x_i x^j\partial_j \, ,\label{eq:specialconformal}\\
    C:& \qquad \qquad \qquad \qquad D&&=t\partial_t + x^i\partial_i \, ,\label{eq:dil} \\
    D^i:& \qquad \qquad \qquad \qquad P_i&&=\partial_i \, ,\\
    E^{ij}:&\qquad \qquad \qquad \qquad L_{ij}&&= x_j\partial_i-x_i\partial_j \, ,\\
    f(x):& \qquad \qquad \qquad \qquad M^{(l,m,n)}&&=x^ly^mz^n \partial_t \, .\label{eq:supertrans}
\end{alignat}
\end{subequations}
Here we have introduced the tuple $(l,m,n)\in \mathbb{Z}^3$ and we have written the spatial coordinates $x \equiv x^1,y\equiv x^2,z\equiv x^3$ explicitly for the generators $M^{(l,m,n)}$. We obtain the generators for spatial special conformal transformations $K_i$, dilatations $D$, translations $P_i$, spatial rotations $L_{ij}$ and `supertranslations' $M^{(l,m,n)}$. Together, they span the infinite extension of the Conformal Carroll Algebra (CCA). Note that the $\{M^{(l,m,n)}\}$ are an infinite set of generators, which contains the Hamiltonian $H \equiv M^{(0,0,0)}$, the boost generators such as $B_1 \equiv M^{(1,0,0)} = x^1 \partial_t$, and the special conformal transformation in the time direction $K_t \equiv -(M^{(2,0,0)}+M^{(0,2,0)}+M^{(0,0,2)}) = -x^jx_j \partial_t$. These three generators form an ideal of the infinitely extended CCA.
\newline

On the other hand, we still have the arbitrary functions $a^{ijI}, b^{iJ}, \dots$ and the constant $G$, which correspond to generators of internal symmetries. The associated generators also span an infinite dimensional symmetry algebra and are related to these quantities as

\begin{subequations}\label{eq:CYMinternal}
\begin{alignat}{3}
 G:& \qquad  \quad &&\qquad \quad \mathcal{I}&&=\epsilon^{ijk} F_{jk} \frac{\partial}{\partial \kappa^i}\, , \label{eq:outlier} \\
  g^{ijk}(t,x):& \quad &&N_{ijk}^{(l,m,n,p)} &&= \frac{1}{4}t^p \bigg[\bigg(\partial_j(x^ly^mz^n) F_{tk} + 2x^ly^mz^n \partial_jF_{tk}\bigg)\frac{\partial}{\partial \kappa^i} \nonumber \\
    & && && \qquad \qquad  \qquad \qquad  \qquad  \qquad \; \; \, + \bigg(\partial_j(x^ly^mz^n) F_{ti}+ 2x^ly^mz^n \partial_jF_{ti}\bigg)\frac{\partial}{\partial \kappa^k}\bigg] \, ,\\
    h^{ij}(t,x) :&    \quad &&Q_{ij}^{(l,m,n,p)}&&=\frac{1}{4}x^ly^mz^n\bigg[\bigg(\partial_t(t^{p}) F_{tj} + 2t^p \partial_t F_{tj}\bigg)\frac{\partial}{\partial \kappa^i} +\bigg(\partial_t(t^{p}) F_{ti} + 2t^p \partial_t F_{ti}\bigg)\frac{\partial}{\partial \kappa^j}\bigg] \, ,  \\
    k^{ij}(t,x):&    \quad &&R_{ij}^{(l,m,n,p)} &&=\frac{1}{2}t^px^ly^mz^n\bigg(F_{ti} \frac{\partial}{\partial \kappa^j} - F_{tj} \frac{\partial}{\partial \kappa^i}\bigg) \, ,\\
  a^{ijI}(x):& \; &&T_{ijI}^{(l,m,n)}&&= 2\bigg[\mathcal{D}_i\bigg(x^ly^mz^nS_I\bigg)\frac{\partial}{\partial \kappa^j}-\mathcal{D}_j\bigg(x^ly^mz^nS_I\bigg)\frac{\partial}{\partial \kappa^i}\bigg]  + \bigg[x^ly^mz^nF_{ij}\bigg] \frac{\partial}{\partial \lambda^I} \, ,\\
         b^{iI}(t,x):& &&U_{iI}^{(l,m,n,p)}&&= t^p x^ly^mz^n\bigg[2\mathcal{D}_t S_I\frac{\partial}{\partial \kappa^i}+F_{ti} \frac{\partial}{\partial \lambda^I}\bigg]\, , \\
         c^{IJ}(x):&  &&V_{IJ}^{(l,m,n)} &&= 4t \partial^i (x^ly^mz^n)\left[S_I, S_J\right]\frac{\partial}{\partial \kappa^i} \nonumber \\
         & && && \qquad \qquad \qquad  + t\bigg[\partial_i\partial^i (x^ly^mz^n) S_J +2\partial^i (x^ly^mz^n) \mathcal{D}_iS_J\bigg]\frac{\partial}{\partial \lambda^I} \nonumber\\
         & && &&\qquad \qquad \qquad - t\bigg[\partial_i\partial^i (x^ly^mz^n) S_I+2\partial^i (x^ly^mz^n) \mathcal{D}_iS_I\bigg] \frac{\partial}{\partial \lambda^J}\nonumber\\
           & && && \qquad \qquad \qquad \qquad \quad   \; \,+\frac{1}{2}x^ly^mz^n\bigg[\lambda_J\frac{\partial}{\partial \lambda^I} -\lambda_I\frac{\partial}{\partial \lambda^J}+S_J\frac{\partial}{\partial S^I} - S_I\frac{\partial}{\partial S^J}\bigg] \, ,\label{eq:Vgenerator}\\
         d^{iIJ}(x):&  &&W_{iIJ}^{(l,m,n)} &&= x^ly^mz^n\left[S_I, S_J\right]\frac{\partial}{\partial \kappa^i} + \frac{1}{4}\bigg[\bigg(\partial_i (x^ly^mz^n)S_J+ 2x^ly^mz^n\mathcal{D}_iS_J\bigg) \frac{\partial}{\partial \lambda^I} \nonumber \\
         & && && \qquad \qquad \qquad \qquad \qquad \qquad \quad \;  \;- \bigg(\partial_i (x^ly^mz^n)S_I+ 2x^ly^mz^n\mathcal{D}_iS_I\bigg) \frac{\partial}{\partial \lambda^J}\bigg] \, ,\\
    n^{IJ}(x):&  &&X_{IJ}^{(l,m,n)}&&=\frac{1}{2}x^ly^mz^n\bigg[S_J \frac{\partial}{\partial \lambda^I}+S_I \frac{\partial}{\partial \lambda^J}\bigg] \, ,\\
        p^{IJ}(t,x):&  &&Y_{IJ}^{(l,m,n,p)}&&=\frac{1}{4}x^ly^mz^n\bigg[\bigg(\partial_t (t^p)\mathcal{D}_t S_J+2t^p \partial_t \mathcal{D}_t S_J\bigg)\frac{\partial}{\partial \lambda^I} \nonumber \\
        & && &&\qquad \qquad \qquad  \qquad \qquad \qquad \qquad \qquad\;\; +\bigg(\partial_t (t^p)\mathcal{D}_t S_I+2t^p \partial_t \mathcal{D}_t S_I\bigg) \frac{\partial}{\partial \lambda^J}\bigg] \, ,\\
        q^{iIJ}(t,x):&  &&Z_{iIJ}^{(l,m,n,p)}&&=\frac{1}{4}t^p\bigg[\bigg(\partial_i (x^ly^mz^n)\mathcal{D}_t S_J+2x^ly^mz^n\partial_i \mathcal{D}_t S_J\bigg)\frac{\partial}{\partial \lambda^I} \nonumber \\
        & && && \qquad \qquad \qquad  \qquad \qquad \qquad +\bigg(\partial_i (x^ly^mz^n)\mathcal{D}_t S_I+2x^ly^mz^n \partial_i \mathcal{D}_t S_I\bigg) \frac{\partial}{\partial \lambda^J}\bigg] \, ,\\
        r^{IJ}(t,x):&  &&J_{IJ}^{(l,m,n,p)} &&= \frac{1}{2}t^p x^ly^mz^n\bigg[\mathcal{D}_t S_J \frac{\partial}{\partial \lambda^I}-\mathcal{D}_t S_I \frac{\partial}{\partial \lambda^J}\bigg]\, .
    \end{alignat}
\end{subequations}
Here we introduced the tuple $(l,m,n,p)\in \mathbb{Z}^4$.
\newline

Next we compute the commutation relations between these generators. First, the generators \eqref{eq:spacetimesymm} have the structure of the CCA 
\begin{subequations}\label{eq:CCAbrackets}
\begin{align}
    [K_i, D] &= -K_i,\qquad \qquad \quad \; \; \; \:
    [K_i, P_j] = 2(L_{ij} - \delta_{ij}D), \qquad \qquad \qquad \;
    [L_{ij}, K_k] = 2K_{[j} \delta_{i]k} \, ,\\
    [D, P_i] &= -P_i, \qquad \qquad 
    [D, M^{(l,m,n)}]=(l+m+n-1)M^{(l,m,n)}, \qquad 
    [P_i, L_{jk}]= 2\delta_{i[k}P_{j]}\, ,\\
     [L_{ij}, L_{kl}] &= 2\delta_{i[k}L_{|j|l]} - 2\delta_{j[k}L_{|i|l]}\, ,\\
    [P_i, M^{(l,m,n)}]&= l M^{(l-1,m,n)}\delta_{ix} + m M^{(l,m-1,n)}\delta_{iy} + n M^{(l,m,n-1)}\delta_{iz}\, , \\ 
    [L_{ij}, M^{(l,m,n)}]&=2l\bigg[M^{(l-1,m+1,n)}\delta_{x[i}\delta_{j]y}+M^{(l-1,m,n+1)}\delta_{x[i}\delta_{j]z}\bigg] \nonumber \\ 
    &\qquad \qquad \qquad \qquad \qquad \qquad \quad   + 2m\bigg[M^{(l+1,m-1,n)}\delta_{y[i}\delta_{j]x}+M^{(l,m-1,n+1)}\delta_{y[i}\delta_{j]z}\bigg]  \\
    &\qquad \qquad \qquad \qquad \qquad \qquad \quad \; + 2n\bigg[M^{(l+1,m,n-1)}\delta_{x[i}\delta_{j]z}+M^{(l,m+1,n-1)}\delta_{y[i}\delta_{j]z}\bigg] \, ,\nonumber \\
    [K_i, M^{(l,m,n)}]&= \bigg[(l+2m+2n-2)M^{(l+1,m,n)}-l\left(M^{(l-1,m+2,n)}+M^{(l-1,m,n+2)} \right)\bigg]\delta_{ix} \nonumber \\
    &\qquad \qquad\; +\bigg[(2l+m+2n-2)M^{(l,m+1,n)}-m\left(M^{(l+2,m-1,n)}+M^{(l,m-1,n+2)}\right)\bigg]\delta_{iy} \\
    &\qquad \qquad\; +\bigg[(2l+2m+n-2)M^{(l,m,n+1)}-n\left(M^{(l+2,m,n-1)}+M^{(l,m+2,n-1)}\right)\bigg]\delta_{iz} \, .\nonumber 
\end{align}
\end{subequations}

For the commutators of the CCA generators with the generators of the internal symmetries, we find that not all commutators close under the full symmetry algebra. For example, if we take $Q_{ij}^{(l,m,n,p)}$ where $p \notin \{0,1\}$, while $l,m,n \in \mathbb{Z}$, we obtain 
\begin{subequations}\label{eq:nonclosure}
\begin{equation}
    \left[D,Q_{ij}^{(l,m,n,p)}\right] = (l+m+n+p)Q_{ij}^{(l,m,n,p)} - \frac{1}{4}pt^{p-1}x^ly^mz^n\bigg[ F_{tj}\frac{\partial}{\partial \kappa^i} +F_{ti}\frac{\partial}{\partial \kappa^j}\bigg] \, .
\end{equation}
Since the last term is clearly not a symmetry of the Carroll $\mathcal{N}=4$ SYM Lagrangian, this bracket does not close under the symmetry Lie algebra. A similar thing happens when $p=1$
\begin{equation}
    \left[D,Q_{ij}^{(l,m,n,1)}\right] = (l+m+n)Q_{ij}^{(l,m,n,1)} + \frac{1}{2}tx^ly^mz^n\bigg[\partial_t F_{tj}\frac{\partial}{\partial \kappa^i} +\partial_t F_{ti}\frac{\partial}{\partial \kappa^j}\bigg]\, .
\end{equation}
However, when $p=0$, the commutator does close
\begin{equation}
    \left[D,Q_{ij}^{(l,m,n,0)}\right] =(l+m+n)Q_{ij}^{(l,m,n,0)} \, .
\end{equation}
\end{subequations}
One may argue that the unwanted extra terms that spoil closure vanish once the constraints \eqref{constraints} are imposed on-shell. Although this is the case in \eqref{eq:nonclosure}, the commutator between the generators $D$ and $T_{ijI}^{(l,m,n)}$ still produces unwanted extra terms that do not vanish on the constraints, i.e. 
\begin{eqnarray}\label{eq:nonclosure_2}
    \left[D, T_{ijI}^{(l,m,n)}\right] = (l+m+n) T_{ijI}^{(l,m,n)} - 4 S_I \partial_{[i} (x^l y^m z^n) \frac{\partial}{\partial \kappa^{j]}} \, . 
\end{eqnarray}

Treating closure as a consistency condition for the generators to be elements of the Lie algebra, we can therefore make a list of all generators which do not close under all brackets with the generators of the CCA
\begin{align}\label{eq:badguys}
    & \left\{Q_{ij}^{(l,m,n,p)},Y_{IJ}^{(l,m,n,p)} \mid p \in \mathbb{Z}_0, \; l,m,n \in \mathbb{Z}\right\} \;\cup  \\
    &\left\{N_{ijk}^{(l,m,n,p)},T_{ijI}^{(l,m,n)},V_{IJ}^{(l,m,n)},W_{iIJ}^{(l,m,n)},Z_{iIJ}^{(l,m,n,p)}\mid l,m,n,p \in \mathbb{Z},\; \exists q \in \{l,m,n,p\}: q \neq 0\right\} \, . \nonumber
\end{align}
By $\exists q \in \{l,m,n,p\}: q \neq 0$ we mean not all of the elements of the tuple $(l,m,n,p)$ or $(l,m,n)$ can be zero, i.e. 
\begin{equation}\label{eq:exceptional}
    \left\{N_{ijk}^{(0,0,0,0)},T_{ijI}^{(0,0,0)},V_{IJ}^{(0,0,0)},W_{iIJ}^{(0,0,0)},Z_{iIJ}^{(0,0,0,0)}\right\} \, ,
\end{equation} 
close under the commutators with the CCA generators; these commutators vanish, as do the commutators of the CCA with $\mathcal{I}$. It is perhaps interesting to note that the commutators with which the generators \eqref{eq:badguys} exhibit this non-closure behavior are precisely those with the conformal generators: $K_i$ \eqref{eq:specialconformal}, $D$ \eqref{eq:dil} and $M^{(l,m,n)}$ \eqref{eq:supertrans} (except for select $(l,m,n) \in \mathbb{Z}^3$, e.g. $M^{(0,0,0)} =\partial_t$). These are precisely the generators of the CCA, except those of the Carroll group.
\newline

The commutators of the remaining generators of the internal symmetries with the CCA generators can then be written down as
\begin{subequations}
\begin{align}
\left[K_a,R_{ij}^{(l,m,n,p)}\right]&=\bigg[(2p+l+2m+2n)R_{ij}^{(l+1,m,n,p)}-l\left(R_{ij}^{(l-1,m+2,n,p)}+R_{ij}^{(l-1,m,n+2,p)}\right)\bigg]\delta_{ax}\nonumber\\
        &\qquad + \bigg[(2p+2l+m+2n)R_{ij}^{(l,m+1,n,p)}- m\left(R_{ij}^{(l+2,m-1,n,p)}+R_{ij}^{(l,m-1,n+2,p)}\right)\bigg]\delta_{ay}\nonumber\\
        &\qquad + \bigg[(2p+2l+2m+n)R_{ij}^{(l,m,n+1,p)}-n\left(R_{ij}^{(l+2,m,n-1,p)}+ R_{ij}^{(l,m+2,n-1,p)}\right)\bigg]\delta_{az} \, ,\\
        \left[D, R_{ij}^{(l,m,n,p)}\right] &=(l+m+n+p)R_{ij}^{(l,m,n,p)} \, ,\\
        \left[P_a, R_{ij}^{(l,m,n,p)}\right] &= l R_{ij}^{(l-1,m,n,p)}\delta_{ax} + m R_{ij}^{(l,m-1,n,p)}\delta_{ay} + nR_{ij}^{(l,m,n-1,p)}\delta_{az} \, ,\\
        \left[L_{ab},R_{ij}^{(l,m,n,p)}\right] &= 2l \bigg[ R_{ij}^{(l-1,m+1,n,p)}\delta_{a[x}\delta_{y]b}+ R_{ij}^{(l-1,m,n+1,p)}\delta_{a[x}\delta_{z]b}\bigg] \nonumber \\
        &\qquad \qquad \qquad \qquad \; \; \; \; + 2m \bigg[R_{ij}^{(l+1,m-1,n,p)}\delta_{a[y}\delta_{x]b}+ R_{ij}^{(l,m-1,n+1,p)}\delta_{a[y}\delta_{z]b}\bigg] \nonumber \\
         &\qquad \qquad \qquad \qquad \; \; \; \; \;+ 2n\bigg[R_{ij}^{(l+1,m,n-1,p)}\delta_{a[z}\delta_{x]b}+ R_{ij}^{(l,m+1,n-1,p)}\delta_{a[z}\delta_{y]b}\bigg]\, ,\\
        \left[M^{(q,r,s)},R_{ij}^{(l,m,n,p)} \right]&= pR_{ij}^{(l+q,m+r,n+s,p-1)} \, .
    \end{align}
\end{subequations}
We can obtain the expressions for all other generators --  $ Q_{ij}^{(l,m,n,0)}, U_{iI}^{(l,m,n,p)},X_{IJ}^{(l,m,n)},Y_{IJ}^{(l,m,n,0)}, J_{IJ}^{(l,m,n,p)}$ -- by simply substituting in these generators and setting $p=0$ where necessary. Finally, there are commutators between internal symmetries and $V_{IJ}^{(l,m,n)}$ \eqref{eq:Vgenerator} due to the derivatives on the scalars $S^I$ and the Lagrange multipliers $\lambda^I$. For completeness, we state them here as well. 
\begin{subequations}
\begin{align}
    \left[V_{IJ}^{(l,m,n)},T_{ijK}^{(q,r,s)}\right] &= T^{(l+q,m+r,n+s)}_{ij[J}\delta_{I]K}\, ,\\
    \left[V_{IJ}^{(l,m,n)},U_{iK}^{(q,r,s,p)}\right] &= U_{i[K}^{(l+q,m+r,n+s,p)}\delta_{I]K}\, ,\\
    \left[V_{IJ}^{(l,m,n)},V_{KL}^{(q,r,s)}\right] &= 2\delta_{[I|[L}V_{K]|J]}^{(l+q,m+r,n+s)} \, ,\\
    \left[V_{IJ}^{(l,m,n)},W_{iKL}^{(q,r,s)}\right] &= 2\delta_{[I|[L|}\bigg[ W_{i|K]|J]}^{(l+q,m+r,n+s)}+ \frac{l}{2} X^{(l+q-1,m+r,n+s)}_{|J]|K]}\delta_{ix} \nonumber \\
    & \qquad \qquad \qquad \qquad \qquad \qquad \qquad \quad \; \; + \frac{m}{2} X^{(l+q,m+r-1,n+s)}_{|J]|K]}\delta_{iy} + \frac{n}{2} X^{(l+q,m+r,n+s-1)}_{|J]|K]}\delta_{iz}\bigg]\, ,\\
    \left[V_{IJ}^{(l,m,n)},X_{KL}^{(q,r,s)}\right] &=  2\delta_{[I|(L}X^{(l+q,m+r,n+s)}_{K)|J]} \, ,\\
    \left[V_{IJ}^{(l,m,n)},Y_{KL}^{(q,r,s,p)}\right]&= 2\delta_{[I|(L}Y^{(l+q,m+r,n+s,p)}_{K)|J]}\, ,\\
    \left[V_{IJ}^{(l,m,n)},Z_{iKL}^{(q,r,s,p)}\right] &= 2\delta_{[I|(L|}\bigg[ Z_{i|K)|J]}^{(l+q,m+r,n+s,p)}+ \frac{l}{2} J^{(l+q-1,m+r,n+s,p)}_{|J)|K]}\delta_{ix} \nonumber \\
    & \qquad \qquad \qquad \qquad \qquad \qquad \qquad \quad + \frac{m}{2} J^{(l+q,m+r-1,n+s,p)}_{|J)|K]}\delta_{iy} + \frac{n}{2} J^{(l+q,m+r,n+s-1,p)}_{|J)|K]}\delta_{iz}\bigg] \, ,\\
    \left[V_{IJ}^{(l,m,n)},J_{KL}^{(q,r,s,p)}\right] &= 2\delta_{[I|[L}J^{(l+q,m+r,n+s,p)}_{K]|J]}\, .
\end{align}
\end{subequations}
Interestingly, the internal symmetries close under their own algebra.
\newline

In summary, the generators under which all the commutators close off-shell are (with  $l,m,n,p \in \mathbb{Z}$)
\begin{align}
  &\bigg\{K_i, D, P_i, L_{ij}, M^{(l,m,n)}\bigg\}\cup \bigg\{\mathcal{I}, N_{ijk}^{(0,0,0,0)},Q_{ij}^{(l,m,n,0)},R_{ij}^{(l,m,n,p)}, \nonumber \\
  & \qquad \qquad \qquad  \qquad \qquad \qquad \qquad T^{(0,0,0)}_{ijI},U_{iI}^{(l,m,n,p)}, V_{IJ}^{(0,0,0)},W_{iIJ}^{(0,0,0)},X_{IJ}^{(l,m,n)},Y_{IJ}^{(l,m,n,0)},Z_{iIJ}^{(0,0,0,0)},J_{IJ}^{(l,m,n,p)}\bigg\} \, .
\end{align}
In addition, the internal generators that close on-shell on the constraints \eqref{constraints} are all the generators listed in \eqref{eq:CYMinternal}, except for 
\begin{equation}
    \left\{T^{(l,m,n)}_{ijI}, V_{IJ}^{(l,m,n)},W_{iIJ}^{(l,m,n)} \mid \exists q \in \{l,m,n\}: q \neq 0\right\}.
\end{equation}

\end{document}